\documentclass[conference,a4paper]{IEEEtran}
\usepackage{xcolor}
\usepackage{balance}
\usepackage{subcaption} %
\usepackage{url}

\usepackage{tikz}
\usetikzlibrary{arrows.meta,positioning,calc}
\ifCLASSINFOpdf
 
\usepackage{amsmath,amssymb}

\ifCLASSOPTIONcompsoc
 \usepackage[caption=false,font=normalsize,labelfont=sf,textfont=sf]{subfig}
\else
 \usepackage[caption=false,font=footnotesize]{subfig}
\fi

\begin{document}
\title{Sampling Reduction for Near-Field to Far-Field Transformation of Bistatic Radar Target Measurements }

\author{\IEEEauthorblockN{
Getahun Kibret Tadesse\IEEEauthorrefmark{1},   % 1st author, 1st affiliations
Thomas Dallmann\IEEEauthorrefmark{2},   % 2nd author, 2nd affiliations
%Author Three\IEEEauthorrefmark{3},    % 3rd author, 3rd affiliations
%Author Four\IEEEauthorrefmark{4}      % 4th author, 4th affiliations
}                                     % ...
%\\

\IEEEauthorblockA{
Radio Technologies for Automated and Connected Vehicles Research Group at
Thüringer Innovationszentrum Mobilität,\\
Technische Universität Ilmenau, Germany\\
getahun-kibret.tadesse@tu-ilmenau.de
}
}

%\IEEEauthorblockA{\IEEEauthorrefmark{3}% 3rd affiliations
%(Affiliation): dept. name of organization, name/acronyms of organization, City, Country, e-mail address*}
%\IEEEauthorblockA{\IEEEauthorrefmark{4}% 4th affiliations
%(Affiliation): dept. name of organization, name/acronyms of organization, City, Country,
 %e-mail address*}  
 %\IEEEauthorblockA{ \emph{*at least one e-mail address should be indicated above} }



\maketitle

%\begin{abstract}
%Accurate bi-static radar cross-section (RCS) characterization requires
%near-field measurements over multiple dimensions, such as transmitter and
%%%%%%%%%%%%receiver positions, target orientation, frequency, and polarization. The
%%%%%%%%%%%resulting near-field sampling requirements can lead to long measurement
%%%%%%%%%%durations for electrically extended radar targets. This work exploits a-priori
%%%%%%%%%information about the target to reduce the number of newly acquired near-field samples.

%%%%%%%%This paper presents a near-field-to-far-field transformation based on an equivalent
%%%%%%%surface-current reconstruction using a-priori information about the target. It is assumed that the prior target and the target under test (TUT) differ by  localized geometrical modifications. Their complex differential near field is reconstructed using Tikhonov-regularized
%%%%%%least squares solved with the least squares QR-factorization (LSQR) algorithm. The
%%%%%reconstructed differential far field is then coherently combined with the
%%%%prior far field contribution to estimate the TUT bi-static RCS.

%%%The method is preliminary evaluated using simulations  of a complex target at \(6~\mathrm{GHz}\). These results indicate that,for the investigated simulation case, prior information can reduce newly acquired sampling locations by \(90\%\) while maintaining
%%close agreement with the CST reference.
%\end{abstract}
\begin{abstract}
Accurate bistatic radar cross-section (RCS) characterization requires
near-field measurements over multiple dimensions, which, under typical
near-field sampling requirements, can lead to long measurement durations for
electrically large radar targets. This paper presents a near-field-to-far-field transformation based on an equivalent surface-current reconstruction using a-priori information about the target to reduce the number of newly acquired near-field samples. It is assumed that the prior target and the target under test (TUT) differ by  localized geometrical modifications. Their complex differential near field is reconstructed using
Tikhonov-regularized least squares QR-factorization (LSQR) method.
The reconstructed differential far-field contribution is coherently combined
with the prior far-field contribution to estimate the bi-static RCS of the
TUT. The method is preliminarily evaluated using  simulations of a complex target at 6~GHz. The results indicate that, for the investigated simulation case, prior information can be used to thin out the sampling grid  by 90\% while maintaining
close agreement with the simulation baseline.
\end{abstract}
\vskip0.5\baselineskip

% \fi
% \IEEEpeerreviewmaketitle

\section{Introduction}

Sixth-generation (6G) wireless systems increasingly integrate communication, sensing, and localization within shared radio-frequency infrastructure. Integrated sensing and communication (ISAC) enables environmental perception using communication networks. In distributed MIMO-ISAC, spatially separated transmitters (TXs) and receivers (RXs) enable bistatic or multistatic sensing, providing diverse target views for characterizing traffic objects and their electromagnetic signatures in realistic environments \cite{Thoma2021JCRS,ThomaDallmann2023DistributedISAC}.

A key quantity for describing the scattering behavior of a target under
multistatic sensing is its bistatic RCS. Accurate
bistatic RCS data is required for target-signature analysis, sensing-system
design, and validation of electromagnetic target models. For electrically
large targets, conventional direct far-field RCS measurements can require
ranges of several hundred metres or kilometres, which may exceed the available
space in typical indoor facilities \cite{Knott2004}. Compact ranges can reduce
the required physical range length by generating an approximate plane-wave
illumination region within an anechoic chamber. However, they require
high-quality reflectors, and edge diffraction limits the size and quality of
the quiet zone. In addition, bistatic compact-range measurements require
additional mechanical arrangements to realize independent transmitter and
receiver geometries \cite{Broquetas1998SphericalWave}.

To overcome these limitations, near-field-to-far-field transformations
(NFFFT) are an established approach for reducing the distance dependence of
radar target measurements. Existing target-oriented NFFFT methods are
predominantly based on monostatic data. Image-based techniques reconstruct an
inverse synthetic aperture radar (ISAR) image from near-field measurements and
derive the far field RCS from the reconstructed image
\cite{Broquetas1998SphericalWave}. Other approaches extrapolate monostatic
near-field data using filtering functions or wavefront-curvature correction
\cite{Chao2013}. Only a limited number of methods address bistatic
NFFFT. These include the use of Kell's equivalence theorem for small bistatic
angles \cite{Kell1965}, transformations based on bistatic near-field data
under fixed plane-wave illumination, and compressed-sensing extensions of
image-based NFFFT algorithms for undersampled measurements
\cite{LaHaie2019,Zhuo2022}.

An NFFFT can also be performed using modal representations, such as
spherical-wave expansion (SWE), or equivalent surface-current reconstruction.
Both approaches are suitable for determining bistatic RCS from measured
near-field data. In SWE, the measured field is represented by a superposition
of analytical modes, whereas equivalent-current reconstruction represents the
field by equivalent currents on a surface enclosing or associated with the
target. The scattered far field, and hence the bistatic RCS, can be evaluated
from either representation at arbitrary observation directions
\cite{Hansen1988,Eibert2009,Eibert2010}.

Although NFFFTs avoid kilometre-scale far-field ranges, they does not remove the
substantial measurement effort required to characterize bistatic scattering
from electrically large targets. In a fully bistatic measurement, both the incident and observation directions
must be sampled. For full-sphere coverage with \(1^\circ\) angular steps,
approximately \(360 \times 181 = 65{,}160\) transmitter positions and the
same number of receiver positions are required. This nominally corresponds to
\((360 \times 181)^2 \approx 4.25 \times 10^9\)
transmitter--receiver configurations at a single frequency and polarization.
Due to reciprocity, interchange of the transmitter and
receiver positions provides the same scattering information, reducing the
number of unique configurations by  a factor of two to
\(2.12 \times 10^9\). Even with an optimistic acquisition time of
\(50\,\mathrm{ms}\) per configuration, this would require approximately
\(1{,}230\) days of measurement time.
The effort increases further when multiple frequencies, polarizations or
target orientations are considered. Thus, reducing the number of newly
acquired near-field samples while preserving bistatic RCS accuracy is
essential.

This work investigates a prior-assisted equivalent-current NFFFT method for
bistatic RCS estimation. The prior target and the TUT are
assumed to differ only through localized geometrical modifications. This is a
realistic assumption when the prior model represents an ideal or nominal target,
whereas the actual object may differ locally due to damage, manufacturing
tolerances, or modeling errors. Such local deviations are expected to produce
predominantly localized differences in the equivalent-current distribution.

This locality provides an advantage over SWE, in
which a local target modification generally affects a global set of modal
coefficients. In the equivalent-current representation, the differential
response is likely to correspond to a localized current correction with lower
effective complexity than the complete target response. This property is
exploited to reconstruct the differential contribution from a reduced set of
newly acquired TUT near-field samples.

The complete prior response is retained, whereas a reduced set of TUT
near-field samples is used to reconstruct the differential scattering
contribution. Under identical measurement conditions, subtraction of the prior
and TUT near-fields suppresses their common response and isolates the
contribution of the modification. The prior and differential
equivalent currents are reconstructed using  LSQR, and
their complex far-field contributions are coherently combined to estimate the
bistatic RCS of the TUT.

This paper presents a preliminary simulation-based investigation of the
approach. The remaining sections of this paper are organized as follows: Section~II describes the theory, Section~III the  simulation
setup, Section~IV the results, and Section~V the conclusions.
\section{theory}
%\section{Prior-Assisted Differential NF--FF Transformation}
\subsection{Differential Near-Field Representation}

Let $\mathbf{E}_{\mathrm{NF}}^{\mathrm{prior}}$ and
$\mathbf{E}_{\mathrm{NF}}^{\mathrm{TUT}}$ denote the complex near fields of
the prior target and TUT. Their differential near field $\mathbf{E}_{\mathrm{NF}}^{\Delta}$ is defined as
\begin{equation}
\mathbf{E}_{\mathrm{NF}}^{\Delta}
=
\mathbf{E}_{\mathrm{NF}}^{\mathrm{TUT}}
-
\mathbf{E}_{\mathrm{NF}}^{\mathrm{prior}},
\label{eq:diff_nf}
\end{equation}
As the targets are assumed to differ only locally and be measured under identical conditions,
the common response is suppressed. Hence, $\mathbf{E}_{\mathrm{NF}}^{\Delta}$ is
dominated by the localized modification and is likely to have lower effective complexity than
the total field.

For reduced TUT measurements, the same subset of \(pN_{\mathrm{dense}}\)
locations is selected from both the TUT and dense prior near-field data. Using
the selection matrix
\(\mathbf{S}_{p}\in\mathbb{R}^{pN_{\mathrm{dense}}\times N_{\mathrm{dense}}}\),
the subsampled differential near field is given by
\begin{equation}
\mathbf{E}_{\mathrm{NF}}^{\Delta,(p)}
=
\mathbf{S}_{p}\mathbf{E}_{\mathrm{NF}}^{\mathrm{TUT,dense}}
-
\mathbf{S}_{p}\mathbf{E}_{\mathrm{NF}}^{\mathrm{prior,dense}},
\label{eq:subsampled_diff_nf}
\end{equation}
The resulting \(\mathbf{E}_{\mathrm{NF}}^{\Delta,(p)}\) is used to reconstruct
the differential equivalent currents. And where \(p\) denotes the percentage of dense near-field sampling locations
retained for the TUT measurements.
\subsection{Electric-Field Measurement Equation}
\subsubsection{Integral Equation}
\begin{figure}[h]
\centering

\begin{tikzpicture}[
    scale=0.60,
    transform shape,
    >=Stealth,
    line cap=round,
    line join=round
]

% ============================================================
% OUTER SURFACE Sigma_M
% ============================================================

\draw[
    blue!65!black,
    line width=1.1pt
]
(0,0) circle (2.45);

% ============================================================
% REGION OMEGA
% ============================================================

\filldraw[
    fill=green!20,
    draw=green!35!black,
    line width=1.1pt
]
(-0.75,1.35)

% upper-left
.. controls (-1.15,1.40)
           and (-1.42,1.18)
           .. (-1.48,0.65)

% left side
.. controls (-1.55,0.10)
           and (-1.40,-0.50)
           .. (-0.90,-0.95)

% lower-left
.. controls (-0.50,-1.30)
           and (-0.05,-1.40)
           .. (0.40,-1.28)

% lower-right
.. controls (0.90,-1.12)
           and (1.45,-0.65)
           .. (1.58,-0.12)

% right side
.. controls (1.18,0.40)
           and (1.43,0.75)
           .. (0.92,0.95)

% upper-right
.. controls (0.45,1.10)
           and (0.35,1.25)
           .. (0.15,1.47)

% upper edge
.. controls (-0.10,1.57)
           and (-0.48,1.43)
           .. (-0.75,1.35)
-- cycle;

% ============================================================
% OMEGA
% ============================================================

\node at (-0.25,0.48)
    {\Large $\Omega$};

% ============================================================
% LEFT-SIDE FIELDS
% ============================================================

\node[anchor=east] at (0.160,-1.88)
    {\large $\vec{\mathbf H}_{+}$};

\node[anchor=east] at (-0.50,-1.882)
    {\large $\vec{\mathbf E}_{+}$};

% ============================================================
% INSIDE FIELDS
% ============================================================

\node at (0.10,-0.04)
    {\large $\vec{\mathbf E}_{-}$};

\node at (0.85,-0.04)
    {\large $\vec{\mathbf H}_{-}$};

% ============================================================
% NORMAL ON TOP OF OMEGA
% ============================================================

\draw[
    ->,
    line width=1.1pt
]
(-0.18,1.5) -- (-0.14,1.95);

\node[anchor=south] at (-0.14,1.95)
    {$\tilde{\mathbf n}$};

% ============================================================
% NORMAL ON Sigma_R
% ============================================================

\draw[
    ->,
    line width=1.1pt
]
(2.150,1.25) -- (2.70,1.65);

\node[anchor=west] at (2.70,2.05)
    {$\tilde{\mathbf n}$};

% ============================================================
% SIGMA_R
% ============================================================

\draw[
    ->,
    line width=0.9pt
]
(1.350,1.15) -- (1.22,0.78);

\node[anchor=west] at (1.35,1.15)
    {$\Sigma_R$};

% ============================================================
% SIGMA_M
% ============================================================

\draw[
    ->,
    line width=0.9pt
]
(-2.40,1.95) -- (-2.0,1.55);

\node[anchor=east] at (-2.40,1.95)
    {$\Sigma_M$};

% ============================================================
% SURFACE CURRENT J
% ============================================================

\draw[
    ->,
    line width=1.1pt
]
(-1.25,-0.65) -- (-0.68,-1.19);

\node[anchor=east] at (-1.15,-0.95)
    {$\vec{\mathbf J}$};

% ============================================================
% MAGNETIZATION M
% ============================================================

\draw[
    ->,
    line width=1.1pt
]
(-0.58,-1.22) -- (0.48,-1.40);

\node[anchor=north] at (0.25,-1.35)
    {$\vec{\mathbf M}$};

\end{tikzpicture}

\caption{Love-equivalence representation on the reconstruction surface.}
\label{fig:geometry}

\end{figure}
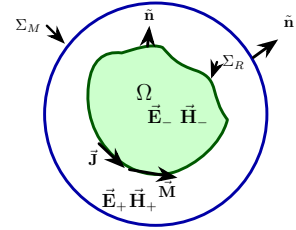

For a PEC target, the reconstruction surface $\Sigma_R$ coincides with the
target surface. Since the tangential total electric field vanishes on this
surface, the magnetic equivalent current is zero. The exterior scattered field
is therefore completely represented by the Love-equivalent electric surface current
\cite{Hansen1988,Eibert2009}:
\vspace{-1.5mm}
\begin{equation}
\mathbf{J}(\mathbf r)
=
\hat{\mathbf n}\times\mathbf{H}_{\mathrm{tot}}(\mathbf r),
\qquad
\mathbf r\in\Sigma_R,
\end{equation}
where $\hat{\mathbf n}$ is the outward unit normal as illustrated in Fig. \ref{fig:geometry}.
Since only the electric field is measured, the unknown electric surface
current is reconstructed through the electric-field radiation operator. The
measured electric field at an observation point $\mathbf r$ is written as
\begin{equation}
\mathbf{E}_{\mathrm{meas}}(\mathbf r)
=
-\eta_0\,
\mathcal{L}\{\mathbf{J}\}(\mathbf r),
\qquad
\mathbf r\in\Sigma_M,
\end{equation}
where $\eta_0$ is the free-space wave impedance, $\Sigma_M$ denotes the
measurement surface, and $\mathcal{L}$ is the electric-field integral
operator,which  is
given by
\begin{equation}
\begin{split}
\mathcal{L}\{\mathbf{J};\mathbf{r}\}
={}& j k_0 \int_{\Sigma_R}
\left[
\mathbf{J}(\mathbf{r}')
+
\frac{1}{k_0^2}
\nabla \left(
\nabla_s' \cdot \mathbf{J}(\mathbf{r}')
\right)
\right] \\
&\times g(\mathbf{r},\mathbf{r}')
\,\mathrm{d}S',
\end{split}
\label{eq:operator_L}
\end{equation}
where $\nabla_s'\cdot$ denotes the surface-divergence operator with respect to
the source point $\mathbf r'$. The free-space Green's function is given by
\begin{equation}
g(\mathbf r,\mathbf r')
=
\frac{
\exp\left(-j k_0
\left|\mathbf r-\mathbf r'\right|\right)
}{
4\pi\left|\mathbf r-\mathbf r'\right|
},
\end{equation}

Here, $k_0$ denotes the free-space wavenumber. The overall sign in the
electric-field equation depends on the adopted time-harmonic convention and
the definition of the Green's function; therefore, this convention is used consistently throughout this paper.
\subsubsection{Method-of-Moments (MoM) Discretization}

The reconstruction surface is meshed with triangular facets, and the electric
surface current is expanded using Rao--Wilton--Glisson (RWG) basis functions $\mathbf{f}$, which avoid edge discontinuities that occur with pulse bases and provide a
divergence-conforming representation of surface currents on arbitrary
triangulated geometries \cite{Eibert2009,Alvarez2007}:
\vspace{-3mm}
\begin{equation}
\mathbf{J}(\mathbf{r})
=
\sum_{n=1}^{N}
I_n\,\mathbf{f}_n(\mathbf{r}),
\end{equation}
where $I_n$ is the unknown coefficient associated with the   $n$th RWG basis function.

\subsubsection{Matrix Inversion}

Substitution of the RWG expansion into the electric-field radiation operator
yields the discrete measurement model 
\begin{equation}
\mathbf{E}_{\mathrm{NF}}
=
\mathbf{Z}\mathbf{I},
\end{equation}
where $\mathbf{Z}$ is assembled from the EFIE using RWG basis functions \cite{Eibert2010,Li2025}.

The current coefficients are found using Tikhonov-regularized least squares,
which stabilizes the ill-posed inverse source problem and suppresses noise
amplification \cite{Tikhonov1977}:
\begin{equation}
\hat{\mathbf{I}}
=
\underset{\mathbf{I}}{\operatorname{arg\,min}}
\left\{
\left\|
\mathbf{Z}\mathbf{I}
-
\mathbf{E}_{\mathrm{NF}}
\right\|_2^2
+
\lambda^2
\left\|
\mathbf{I}
\right\|_2^2
\right\},
\label{eq:tikhonov}
\end{equation}
where $\lambda$ is the regularization parameter. 
%\begin{equation}
%\begin{bmatrix}
%\mathbf{Z}\\
%\lambda\mathbf{I}
%\end{bmatrix}
%\hat{\mathbf{I}}
%=
%\begin{bmatrix}
%\mathbf{E}_{\mathrm{NF}}\\
%\mathbf{0}
%\end{bmatrix}.
%\label{eq:augmented_system}
%\end{equation}

The same procedure reconstructs the prior current
$\hat{\mathbf{I}}^{\mathrm{prior}}$ and the differential current
$\hat{\mathbf{I}}^{\Delta}$,
where $\hat{\mathbf{I}}^{\Delta}$ is defined analogously to
$\mathbf{E}_{\mathrm{NF}}^{\Delta}$ in \eqref{eq:diff_nf}. 
\subsection{Far-Field Reconstruction}

The far field radiated by the reconstructed equivalent current is obtained
using the free-space radiation operator \cite{Eibert2009,Eibert2010,Alvarez2009}:
\begin{equation}
\hat{\mathbf{E}}_{\mathrm{FF}}^{\mathrm{prior}}
=
\mathcal{F}\!\left\{\hat{\mathbf{J}}^{\mathrm{prior}}\right\},
\qquad
\hat{\mathbf{E}}_{\mathrm{FF}}^{\Delta}
=
\mathcal{F}\!\left\{\hat{\mathbf{J}}^{\Delta}\right\},
\label{eq:far}
\end{equation}
The electric far-field in the direction  $\hat{\mathbf r}$ which is based on the electric surface-current distribution $\mathbf{J}(\mathbf r')$, is defined as:
\begin{equation}
\begin{split}
\mathbf{E}_{\mathrm{FF}}(\hat{\mathbf{r}})
={}& \mathcal{F}\{\mathbf{J}\} \\
={}& \frac{j k \eta e^{-jkr}}{4\pi r}
\hat{\mathbf{r}} \times
\left[
\hat{\mathbf{r}} \times
\int_{\Sigma_R}
\mathbf{J}(\mathbf{r}')
e^{j k \hat{\mathbf{r}} \cdot \mathbf{r}'}
\,\mathrm{d}S'
\right],
\end{split}
\label{eq:ff_integral}
\end{equation}
where $k$ and $\eta$ are the wavenumber and intrinsic impedance of the
background medium, respectively.

\subsection{Prior-Assisted Far-Field Reconstruction}

The far field of the TUT is estimated by coherent superposition of the prior
and differential far-field contributions:
\begin{equation}
\hat{\mathbf{E}}_{\mathrm{FF}}^{\mathrm{TUT}}
=
\hat{\mathbf{E}}_{\mathrm{FF}}^{\mathrm{prior}}
+
\hat{\mathbf{E}}_{\mathrm{FF}}^{\Delta},
\label{eq:ff_sum}
\end{equation}

Thus, the dense prior data represent the common response of the target, while
the sparse TUT near-field data is used to reconstruct only the correction
associated with the localized geometrical modification.
\subsection{Bistatic RCS Evaluation}

The bistatic RCS is calculated from the reconstructed total transverse
scattered far field as \cite{Knott2004}:
\begin{equation}
\hat{\sigma}_{\mathrm{B}}(\theta,\phi)
=
4\pi R^2
\frac{
\left\lvert
\hat{\mathbf{E}}_{\mathrm{FF}}^{\mathrm{TUT}}(\theta,\phi;R)
\right\rvert^2
}{
\left\lvert
\mathbf{E}_{\mathrm{inc}}
\right\rvert^2
},
\label{eq:bistatic_rcs}
\end{equation}
where $\hat{\mathbf{E}}_{\mathrm{FF}}^{\mathrm{TUT}}(\theta,\phi;R)$ is
the reconstructed scattered electric-field vector evaluated at the
far-field reference distance $R$, and $\mathbf{E}_{\mathrm{inc}}$ is the
incident electric field at the target reference point. 
\subsection{Overview of the Proposed Reconstruction Procedure}

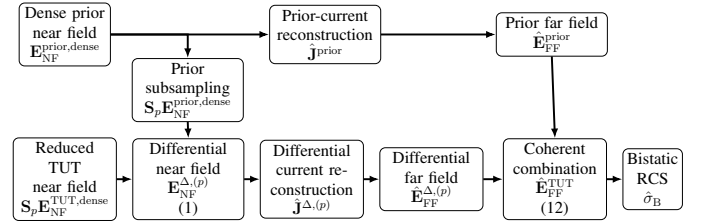
\begin{figure}[h]
\centering
\resizebox{\columnwidth}{!}{%
\begin{tikzpicture}[
    node distance=3mm and 3mm,
    box/.style={
        draw,
        rounded corners,
        align=center,
        text width=12mm,
        minimum height=8mm,
        inner sep=1mm,
        font=\large
    },
    arrow/.style={
        -{Latex[length=2.5mm]},
        line width=1.4pt
    }
]

% Inputs
\node[box, text width=2.1cm, align=center] (priornf) {Dense prior near field\\
$\mathbf{E}_{\mathrm{NF}}^{\mathrm{prior,dense}}$};

\node[box, below=18mm of priornf,text width=2.4cm, align=center] (tutnf)
{Reduced TUT near field\\
$\mathbf{S}_{p}\mathbf{E}_{\mathrm{NF}}^{\mathrm{TUT,dense}}$};

% Prior branch
\node[box, right=40mm of priornf, text width=2.6cm, align=center] (priorj)
{Prior-current\\ reconstruction\\
$\hat{\mathbf{J}}^{\mathrm{prior}}$};

\node[box, right=29mm of priorj, text width=2.6cm, align=center] (priorff)
{Prior far field\\
$\hat{\mathbf{E}}_{\mathrm{FF}}^{\mathrm{prior}}$};

% Prior subsampling branch
\node[box, below=-2mm of priornf,
      xshift=31mm,text width=2.6cm, align=center] (subsample)
{Prior subsampling\\
$\mathbf{S}_{p}\mathbf{E}_{\mathrm{NF}}^{\mathrm{prior,dense}}$};

% Differential branch
\node[box, right=4mm of tutnf,text width=2.6cm, align=center] (diffnf)
{Differential near field\\
$\mathbf{E}_{\mathrm{NF}}^{\Delta,(p)}$\\
\eqref{eq:diff_nf}};

\node[box, right=4mm of diffnf,text width=2.4cm, align=center] (diffj)
{Differential current reconstruction\\
$\hat{\mathbf{J}}^{\Delta,(p)}$};

\node[box, right=4mm of diffj,text width=2.4cm, align=center] (diffff)
{Differential far field\\
$\hat{\mathbf{E}}_{\mathrm{FF}}^{\Delta,(p)}$};

% Output branch
\node[box, right=5mm of diffff,text width=2.4cm, align=center] (combine)
{Coherent combination\\
$\hat{\mathbf{E}}_{\mathrm{FF}}^{\mathrm{TUT}}$\\
\eqref{eq:ff_sum}};

\node[box, right=4mm of combine,text width=1.30cm, align=center] (rcs)
{Bistatic RCS\\
$\hat{\sigma}_{\mathrm{B}}$};

% Prior-current branch
\draw[arrow] (priornf) -- (priorj);
\draw[arrow] (priorj) -- (priorff);

% Differential-current branch
\draw[arrow] (tutnf) -- (diffnf);
\draw[arrow] (diffnf) -- (diffj);
\draw[arrow] (diffj) -- (diffff);

% Prior field used for differential field
%\draw[arrow] (priornf.east) -| (diffnf.north);
% Prior subsampling path
\draw[arrow] (priornf.east) -| (subsample.north);
\draw[arrow] (subsample.south) -- (diffnf.north);
% Coherent far-field combination
\draw[arrow] (priorff.south) -- (combine.north);
\draw[arrow] (diffff) -- (combine);
\draw[arrow] (combine) -- (rcs);

\end{tikzpicture}%
}
\caption{Workflow of the proposed prior-assisted equivalent-
current NFFFT method.}
\label{fig:workflow}
\end{figure}

Fig.~\ref{fig:workflow} summarizes the proposed method. 
\vspace{-1mm}
\section{Simulation Setup}
The proposed method is evaluated in CST Studio Suite using simulated complex
electric fields sampled on a spherical near-field surface.
\subsection{Target Configuration}
\begin{figure}[!t]
\centering
% ================= LEFT: TUT =================
\begin{minipage}[b]{0.450\columnwidth}
    \centering
    \begin{tikzpicture}[
        x=\linewidth,
        y=\linewidth,
        >=Latex
    ]
        % Background image
        \node[inner sep=0, anchor=south west] (img) at (0,0)
        {\includegraphics[
            width=\linewidth,
            trim={0 0 20 0},
            clip
        ]{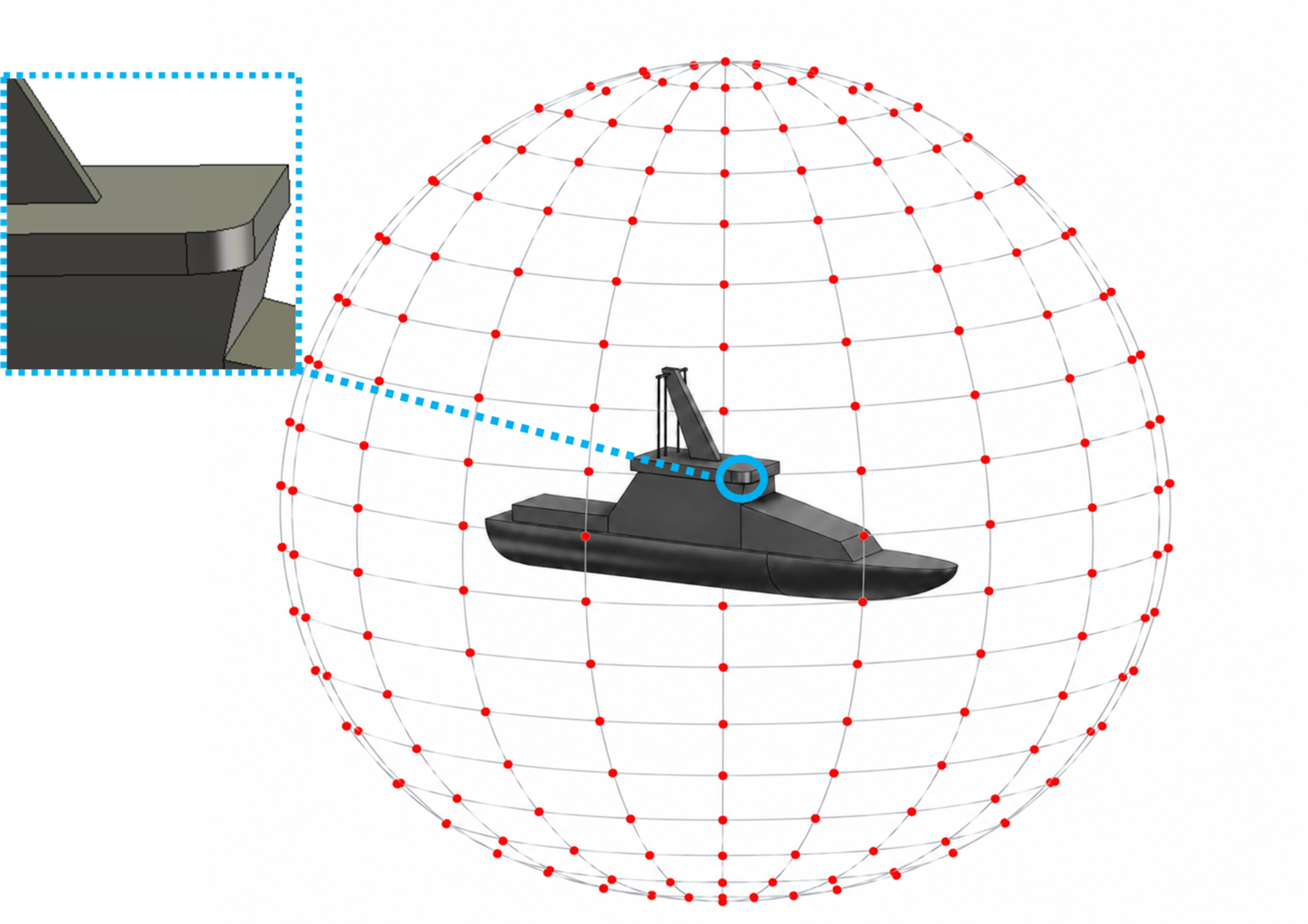}};
% Coordinate origin
\coordinate (O) at
    ([xshift=0.57\linewidth,
      yshift=0.32\linewidth]img.south west);

% z-axis
\draw[blue, line width=0.6pt, -{Latex[length=1.5mm]}]
    (O) -- ++(0,0.45\linewidth)
    node[above, font=\normalsize\bfseries] {$z$};

% x-axis
\draw[red,line width=0.6pt, -{Latex[length=1.5mm]}]
    (O) -- ++(0.41\linewidth,-0.05\linewidth)
    node[right, xshift=-1mm, yshift=-1mm, font=\normalsize\bfseries] {$x$};

% y-axis
\draw[green!50!black, line width=0.6pt, -{Latex[length=1.5mm]}]
    (O) -- ++(0.37\linewidth,0.15\linewidth)
    node[above right, xshift=-1mm, yshift=-2mm, font=\normalsize\bfseries] {$y$};
           % Theta angle
\draw[black!75, line width=0.5pt, -{Latex[length=1.5mm]}]
    ($(img.south west)+(0.57\linewidth,0.73\linewidth)$)
    arc[
        start angle=90,
        end angle=45,
        radius=0.11\linewidth
    ];

\node[font=\normalsize\bfseries]
    at ($(img.south west)+(0.62\linewidth,0.79\linewidth)$)
    {$\theta$};

% ================= PHI ANGLE (REVERSED) =================
\draw[black!75, thick, -{Latex[length=2mm]}]
    ($(img.south west)+(0.95\linewidth,0.28\linewidth)$)
    arc[
        start angle=0,
        end angle=40,
        radius=0.15\linewidth
    ];

\node[font=\normalsize\bfseries]
    at ($(img.south west)+(0.97\linewidth,0.40\linewidth)$)
    {$\phi$};
    \end{tikzpicture}

    \small (a) TUT
\end{minipage}
\hfill
%\hspace{-0.1mm}
% ================= RIGHT: PRIOR TARGET =================
\begin{minipage}[b]{0.450\columnwidth}
    \centering
    \begin{tikzpicture}[
        x=\linewidth,
        y=\linewidth,
        >=Latex
    ]
        % Background image
        \node[inner sep=0, anchor=south west] (img) at (0,0)
        {\includegraphics[
            width=\linewidth,
            trim={0 0 0 0},
            clip
        ]{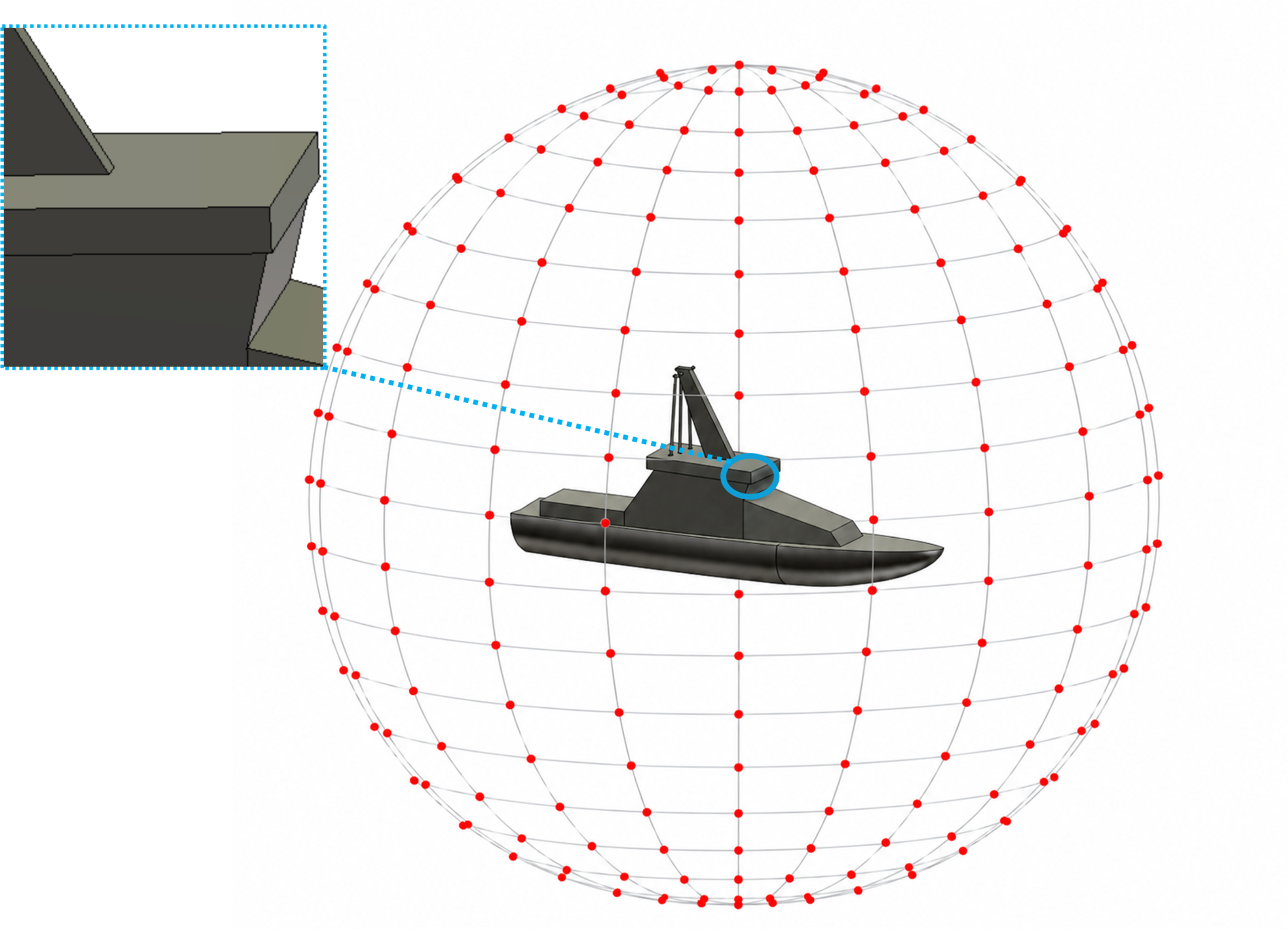}};
% Coordinate origin
\coordinate (O) at
    ([xshift=0.57\linewidth,
      yshift=0.32\linewidth]img.south west);

% z-axis
\draw[blue, line width=0.6pt, -{Latex[length=1.5mm]}]
    (O) -- ++(0,0.42\linewidth)
    node[above, font=\normalsize\bfseries] {$z$};

% x-axis
\draw[red,line width=0.6pt, -{Latex[length=1.25mm]}]
    (O) -- ++(0.39\linewidth,-0.05\linewidth)
    node[right,xshift=-1mm, yshift=-1mm ,font=\normalsize\bfseries] {$x$};

% y-axis
\draw[green!50!black, line width=0.6pt, -{Latex[length=1.5mm]}]
    (O) -- ++(0.37\linewidth,0.14\linewidth)
    node[above right, xshift=-1mm, yshift=-2mm, font=\normalsize\bfseries] {$y$};
           % Theta angle
\draw[black!75, line width=0.5pt, -{Latex[length=1.5mm]}]
    ($(img.south west)+(0.58\linewidth,0.70\linewidth)$)
    arc[
        start angle=90,
        end angle=45,
        radius=0.11\linewidth
    ];

\node[font=\normalsize\bfseries]
    at ($(img.south west)+(0.64\linewidth,0.78\linewidth)$)
    {$\theta$};

% ================= PHI ANGLE (REVERSED) =================
\draw[black!75, thick, -{Latex[length=2mm]}]
    ($(img.south west)+(0.95\linewidth,0.28\linewidth)$)
    arc[
        start angle=0,
        end angle=45,
        radius=0.15\linewidth
    ];

\node[font=\normalsize\bfseries]
    at ($(img.south west)+(0.98\linewidth,0.40\linewidth)$)
    {$\phi$};
    \end{tikzpicture}
    \small (b) Prior target
\end{minipage}
\caption{Target configurations used in the simulation.}
\label{fig:target_configurations}
\end{figure}
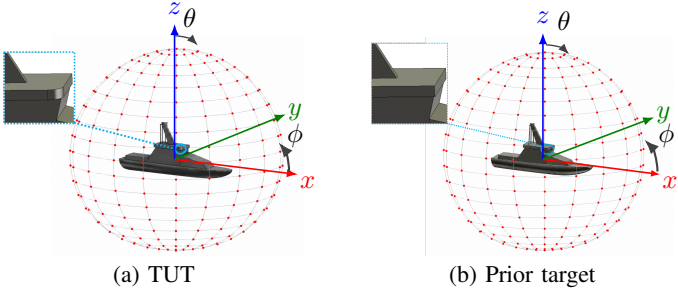

Fig.~\ref{fig:target_configurations} compares the nominal IHF boat CAD model,
used as prior information, with the manufactured TUT
\cite{Dallmann2014}. The TUT differs from the nominal model through small manufacturing-induced
geometric deformations. Zoomed-in views of both targets highlight these local
differences.

\subsection{Simulated Near-Field Configuration}
Fig.~\ref{fig:target_configurations} also defines, the target-fixed Cartesian
coordinate system with the  \(x\)-axis along the target length, the \(y\)-axis along
the transverse direction, and the \(z\)-axis vertically upward. A vertically-polarized
plane wave propagates along the negative \(x\)-axis with
\((\theta_{\mathrm{inc}},\phi_{\mathrm{inc}})=(90^{\circ},0^{\circ})\). The complex scattered field is then sampled on a spherical surface centered on the
target using 7700
measurement locations. The same configuration is used for the prior target and
the TUT. Table~\ref{tab:simulation_parameters} lists all the parameters.
\begin{table}[h]
\caption{Simulation and virtual measurement parameters.}
\label{tab:simulation_parameters}
\centering
\begin{tabular}{l l}
\hline
Parameter & Value \\
\hline
Operating frequency & 6 GHz \\
Measurement-surface geometry & sphere radius 1300 mm \\
Observation locations & $N_{\theta}N_{\phi}=70\times110=7700\ $\\
CST field components & $E_x$, $E_y$, $E_z$ \\
Reconstruction components & $E_\theta$, $E_\phi$ \\
Reconstruction data entries & $2\times7700=15400$ complex values \\
\hline
\end{tabular}
\end{table}

\subsection{Reference and Reduced-Data Sampling Strategy}
Following the procedure in Section~II-F, The aforementioned \(N_{\mathrm{dense}}=7700\)  samples are generated and used as dense near-field data for the prior target and the TUT.The dense differential near field $\mathbf{E}^{\Delta,\mathrm{ref}}_{\mathrm{NF}}$ 
is obtained according to\eqref{eq:diff_nf}.
The RCS reconstructed using all TUT samples serves as the full-data reference,
while the direct simulation far-field RCS is used as the baseline.

For the reduced-data case, the complete prior data set is retained, whereas
only \(p=10\%\), corresponding to 770 TUT locations, is used. The differential
field is formed from the corresponding prior and TUT samples according to
\eqref{eq:subsampled_diff_nf}. The bistatic RCS reconstructed from the
resulting differential field is compared with the direct simulation baseline far-field RCS.
\section{Results and Discussion}

\begin{figure}[h]
    \centering
    \begin{minipage}[b]{0.45\columnwidth}
        \centering
        \includegraphics[
            width=\columnwidth,
            trim={0 235 10 210},
            clip
        ]{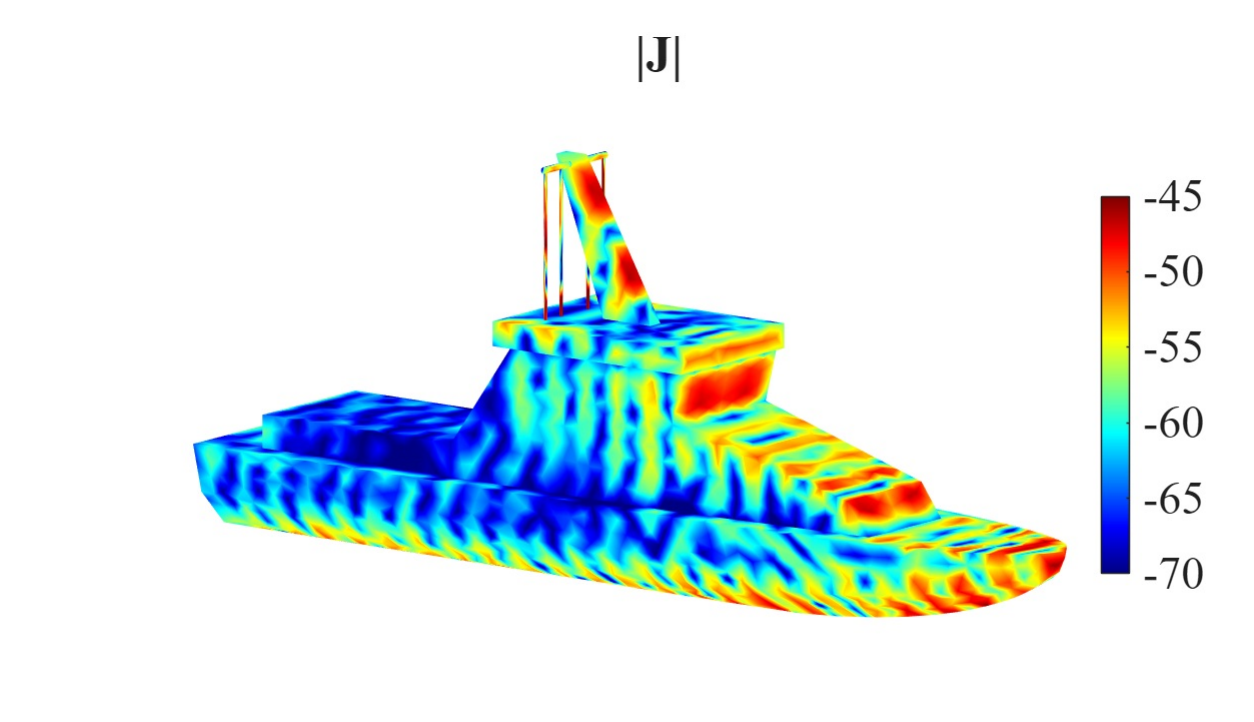}
        \vspace{-1.5mm}
        \centerline{\footnotesize (a) Reconstructed current}
    \end{minipage}\hfill
    \begin{minipage}[b]{0.45\columnwidth}
        \centering
        \includegraphics[
            width=0.8\columnwidth,
            trim={50 40 130 0},
            clip
        ]{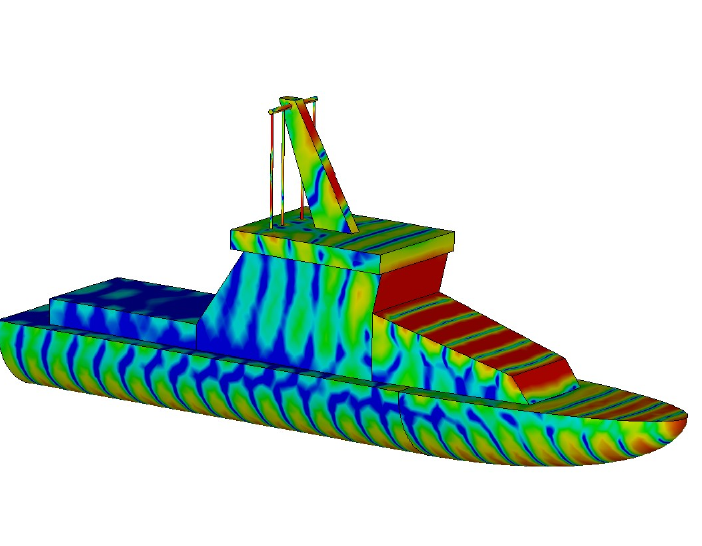}
        \vspace{-1.5mm}
        \centerline{\footnotesize (b) CST current}
    \end{minipage}
    \caption{Surface-current magnitude from dense near-field data:
    (a) reconstructed equivalent current and (b) CST surface current.}
    \label{fig:surface_current_comparison}
\end{figure}

The reconstruction is validated using all \(7700\) near-field samples.
Fig.~\ref{fig:surface_current_comparison} compares the reconstructed
equivalent current with the baseline current projected onto the prior-target
model. The reconstruction captures the main current distribution and the
dominant high-current regions, although local differences remain.
Despite the local differences visible in
Fig.~\ref{fig:surface_current_comparison}, Fig.~\ref{fig:rcs_dense_validation}
shows excellent agreement between the estimated bistatic RCS from the
reconstructed currents and the direct simulation baseline far-field RCS. The agreement over the common azimuth grid is quantified using the root-mean-square error (RMSE):
\begin{equation}
\mathrm{RMSE}
=
\sqrt{
\frac{1}{N_{\phi}}
\sum_{i=1}^{N_{\phi}}
\left(
\hat{\sigma}_{\mathrm{B},i}^{\mathrm{dBsm}}
-
\sigma_{\mathrm{B,CST},i}^{\mathrm{dBsm}}
\right)^2
},
\label{eq:rmse}
\end{equation}
where \(\hat{\sigma}_{\mathrm{B},i}^{\mathrm{dBsm}}\) and \(\sigma_{\mathrm{B,CST},i}^{\mathrm{dBsm}}\) denote the reconstructed and direct-simulation baseline far-field RCS values, respectively, at the \(i\)th azimuth sample. The dense-data reconstruction achieves a low RMSE of \(0.0516\)~dB, indicating that the reconstructed currents preserve the target's relevant scattering characteristics and enable accurate far-field prediction from near-field measurements. The small residual error arises because the far-field is calculated using a finite reconstructed equivalent-current representation rather than directly from the simulation.

\subsection{Reduced-Data Bistatic RCS Reconstruction}
\begin{figure}[!t]
\centering

\begin{subfigure}{\columnwidth}
    \centering
    \includegraphics[width=\columnwidth]{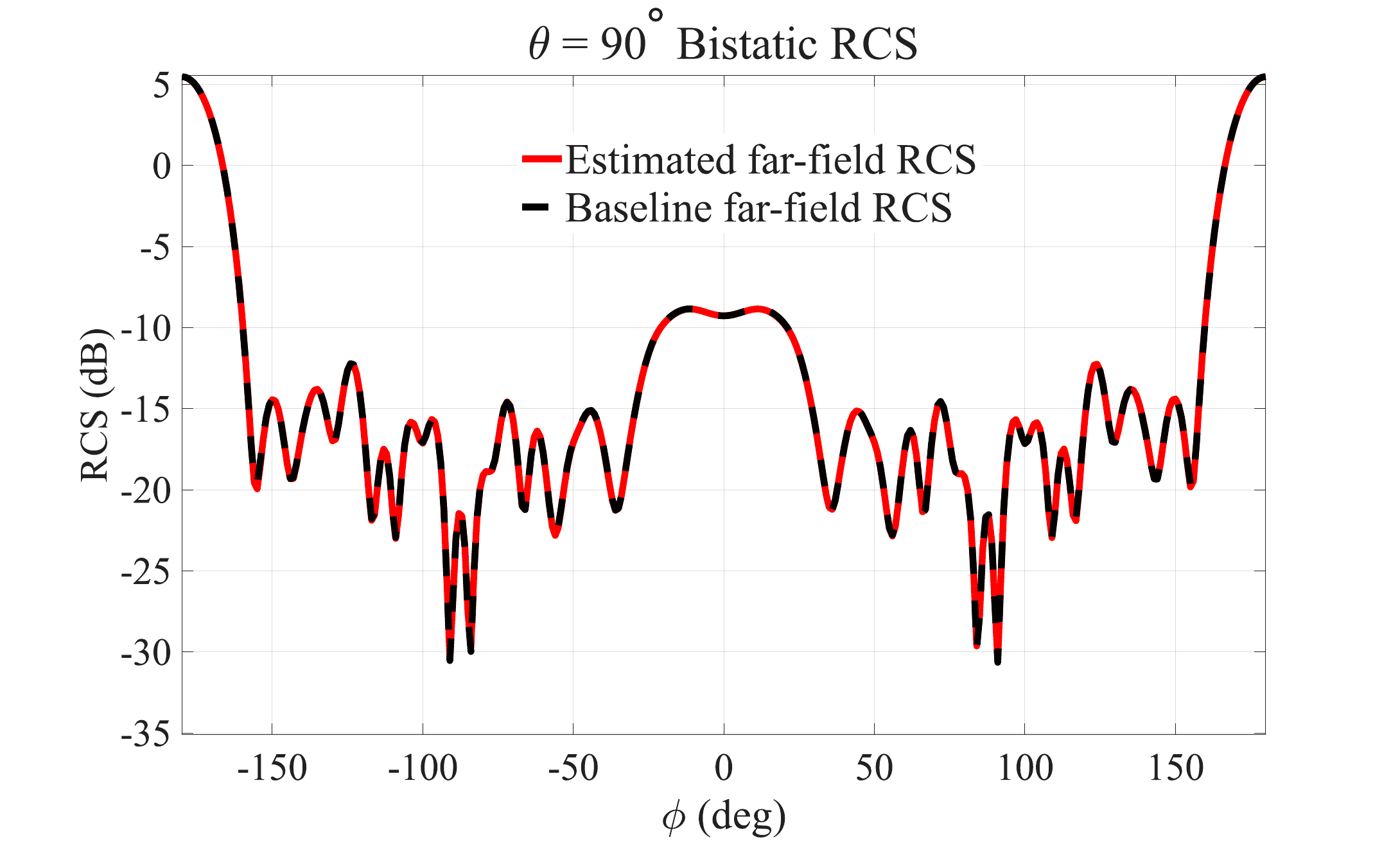}
    \caption{Bistatic RCS on the $\theta=90^\circ$ cut obtained from the dense
reconstruction and compared with the direct baseline far-field RCS .}
    \label{fig:rcs_dense_validation}
\end{subfigure}

\vspace{1mm}

\begin{subfigure}{\columnwidth}
    \centering
    \includegraphics[width=\columnwidth]{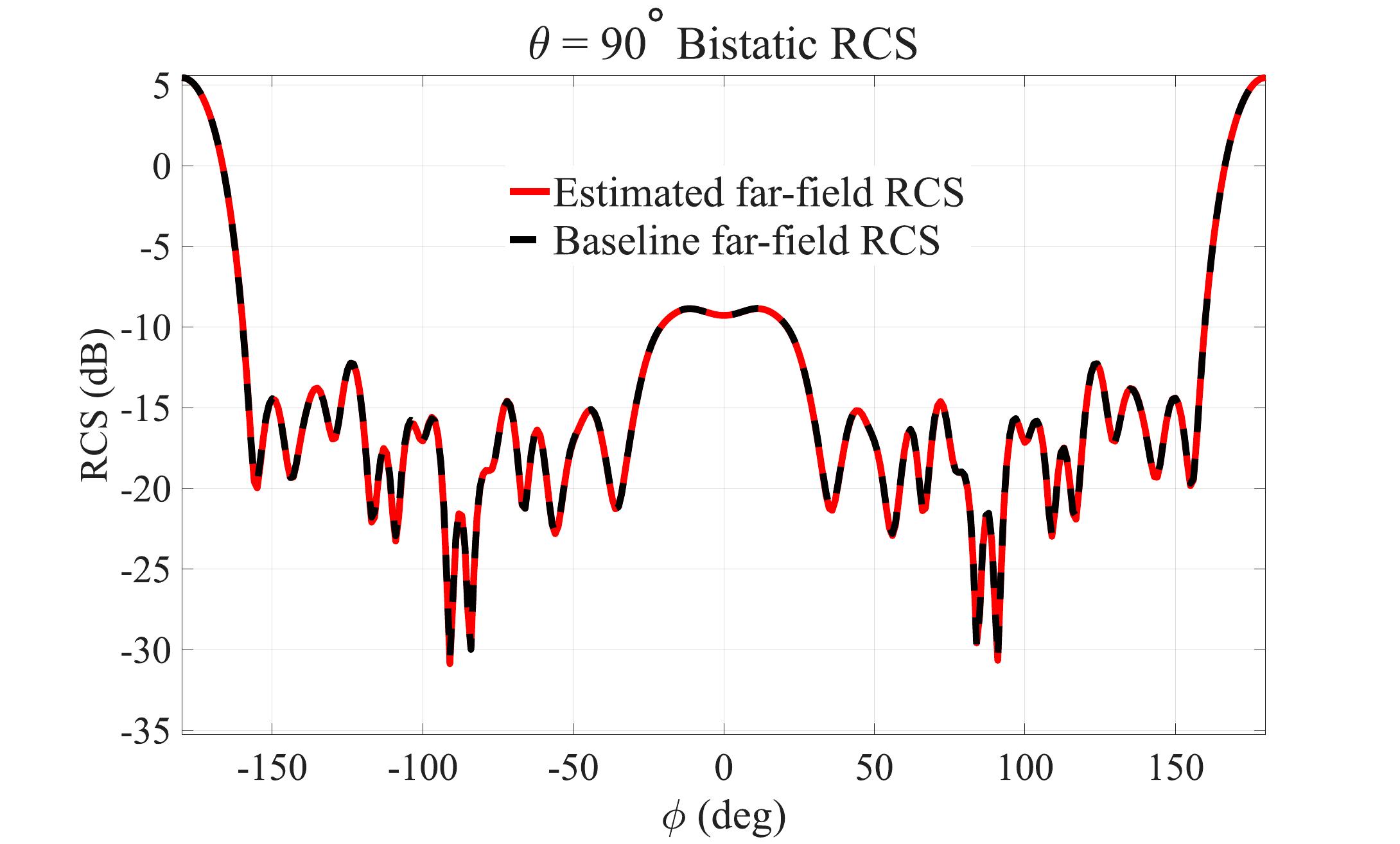}
    \caption{Bistatic RCS reconstructed using \(10\%\) of the TUT near-field data and the
dense prior near-field data, compared with the baseline far-field RCS.}
    \label{fig:reduced_rcs}
\end{subfigure}

\caption{Bistatic RCS as a function of the azimuth angle \(\phi\) for the
\(\theta=90^\circ\) cut.}
\label{fig:rcs_comparison}
\end{figure}

The dense prior near-field data set is retained, while only a subset of TUT
samples is used to reconstruct the differential current. Fig.~\ref{fig:reduced_rcs} compares the bistatic RCS obtained by coherently
combining the prior far-field with the differential far-field reconstructed
from reduced TUT near-field samples against the direct simulation baseline
far-field RCS. The proposed reconstruction follows the simulation baseline curve over the common azimuth grids.This agreement is further quantified using \eqref{eq:rmse}, yielding an RMSE of $0.0770$~dB. Although this value is slightly higher than that obtained in the dense-sampling case, it remains very low, indicating that the proposed method maintains high reconstruction accuracy despite the substantial reduction in measurement samples. Overall, the results demonstrate that the proposed method preserves the bistatic RCS characteristics while reducing the number of newly acquired TUT measurement locations by $90\%$ for the investigated simulation scenario.
Table~\ref{tab:rcs_errors} summarizes the RCS errors defined in \eqref{eq:rmse} for different sub-samplings. 
\begin{table}[h]
\caption{Bistatic RCS errors relative to the CST far-field reference.}
\label{tab:rcs_errors}
\centering
\renewcommand{\arraystretch}{1.1}
\begin{tabular}{c c c c c}
\hline
TUT data & Samples  & RMSE (dB) \\
\hline
Dense ($100\%$) & 7700 & 0.051605  \\
Reduced ($50\%$) & 3850  & 0.051643  \\
Reduced ($10\%$) & 770  & 0.077003  \\
\hline
\end{tabular}\\[-1mm]
%\footnotesize All error values are in dB.
\end{table}
\subsection{Differential Equivalent Surface-Current Reconstruction}
\begin{figure}[h]
\centering
\includegraphics[
    width=0.7\columnwidth,
    trim={0 286 0 232},
    clip
]{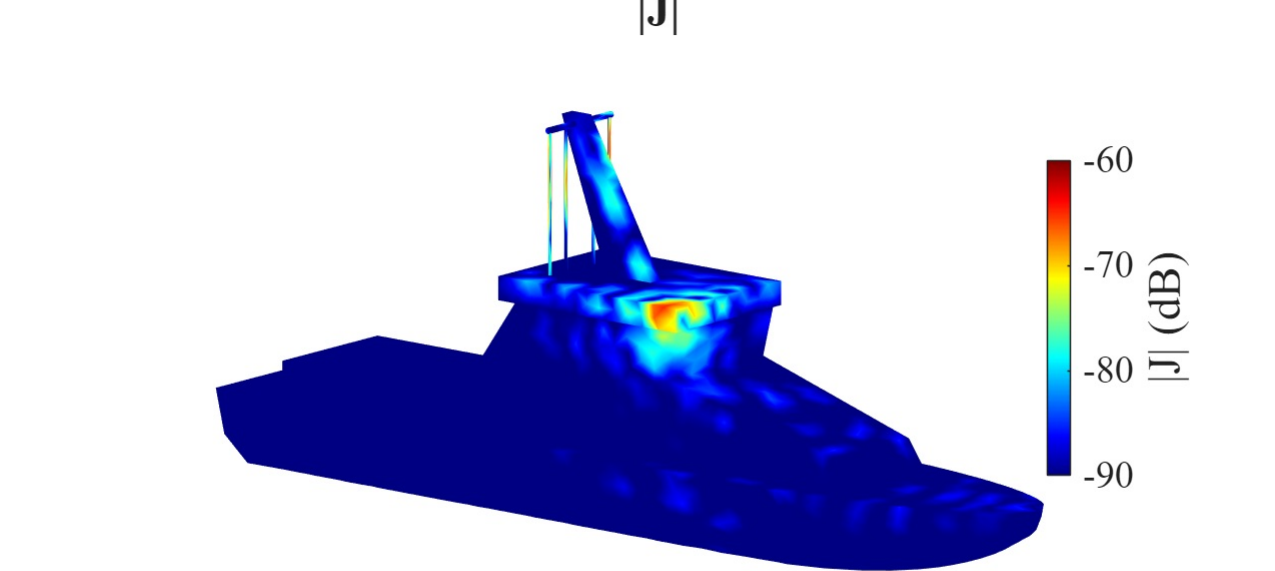}
\caption{Differential equivalent-current magnitude.}
\label{fig:differential_current}
\end{figure}

The dense differential near field in \eqref{eq:diff_nf} is used to reconstruct
the differential equivalent current. Fig.~\ref{fig:differential_current} shows
the magnitude of the reconstructed differential equivalent current. The
dominant response is concentrated around the deformation because subtracting
the prior and TUT fields suppresses the common scattering contributions from
unchanged regions. Weaker responses at the rods and mast, approximately
\(9\)--\(12\)~dB below the peak, indicate coupling between the deformation and
nearby structures. Over most of the target, the differential current is more
than \(20\)~dB below the dominant response. This confirms that the
differential formulation isolates the primary effect of the target modification
while retaining weaker secondary effects. The localized differential response
is subsequently reconstructed using sparse TUT near-field samples.
\vspace{-6mm}
\vspace{-6mm}
\section{Conclusion}

This work presented an initial simulation study of a prior-assisted
equivalent-current NFFFT method for bistatic RCS estimation. The method
reconstructs the differential near-field between a nominal CAD prior and a
modified TUT, and combines the resulting differential response with the prior
far-field.

For the investigated IHF-boat simulation at $6~\mathrm{GHz}$, the dense
reconstruction using all $7700$ TUT locations achieved an RMSE of
$0.0516$~dB. Using only $770$ locations ($10\%$), the RMSE was
$0.0770$~dB, corresponding to a $90\%$ reduction in newly acquired TUT
samples.

This work is a preliminary simulation-based investigation of the proposed
method. Experimental validation is required to determine its suitability for
practical bistatic RCS measurements. In a measurement scenario, the
differential near field may be affected by target misalignment, calibration and
phase-reference errors, measurement noise, transmitter--receiver coupling, and
deviations from the assumed illumination field. The current formulation also
assumes plane-wave illumination, which is restrictive when the transmitter is
located in the near-field region of the TUT. Future work will therefore
validate the method experimentally, assess its robustness to these measurement
uncertainties, and extend it to account for near-field transmitter
illumination.

\section*{Acknowledgment}

This work was carried out within the BiRaUM project and was funded by the
Free State of Thuringia and the European Social Fund Plus under grant 2025 FGR 0020.

\end{document}